\documentclass[
  ,draft            
  ]
  {aipproc}

\layoutstyle{8x11single}

\usepackage{amssymb}     
\usepackage{marvosym}    
\usepackage{upgreek}     

\newcommand{\thalfzero}   {${T^{0\nu}_{1/2}}$}
\newcommand{\thalftwo}    {${T^{2\nu}_{1/2}}$}

\newcommand{\GERDA}       {\mbox{\textsc{Gerda}}}  
\begin{document}

\title{The search for 0$\nu\beta\beta$ decay with the \GERDA\,experiment: status and prospects}


\author{B. Majorovits for the \GERDA\,collaboration}{
  address={Max Planck Institute for Physics, Munich, Germany}
}


\begin{abstract}
 The \GERDA\, experiment is designed to search for neutrinoless double beta decay of $^{76}$Ge using HPGe detectors directly immersed into liquid argon. In its first phase the \GERDA\, experiment has yielded a half life limit on this decay of \thalfzero\,$>$ 2.1$\cdot$10$^{25}$\,yr. A background model has been developed. It explains the measured spectrum well, taking into account only components with distances to the detectors less then 2\,cm. Competitive limits on Majoron accompanied double beta decay have been derived. Phase II of the experiment, now with additional liquid argon veto installed, is presently starting its commissioning phase. First commissioning spectra from calibration measurements are shown, proving that the liquid argon veto leads to a significant reduction of background events. 
\end{abstract}

\maketitle


\section{Introduction}

\begin{figure}[b!]
\includegraphics[height=.2\textheight]{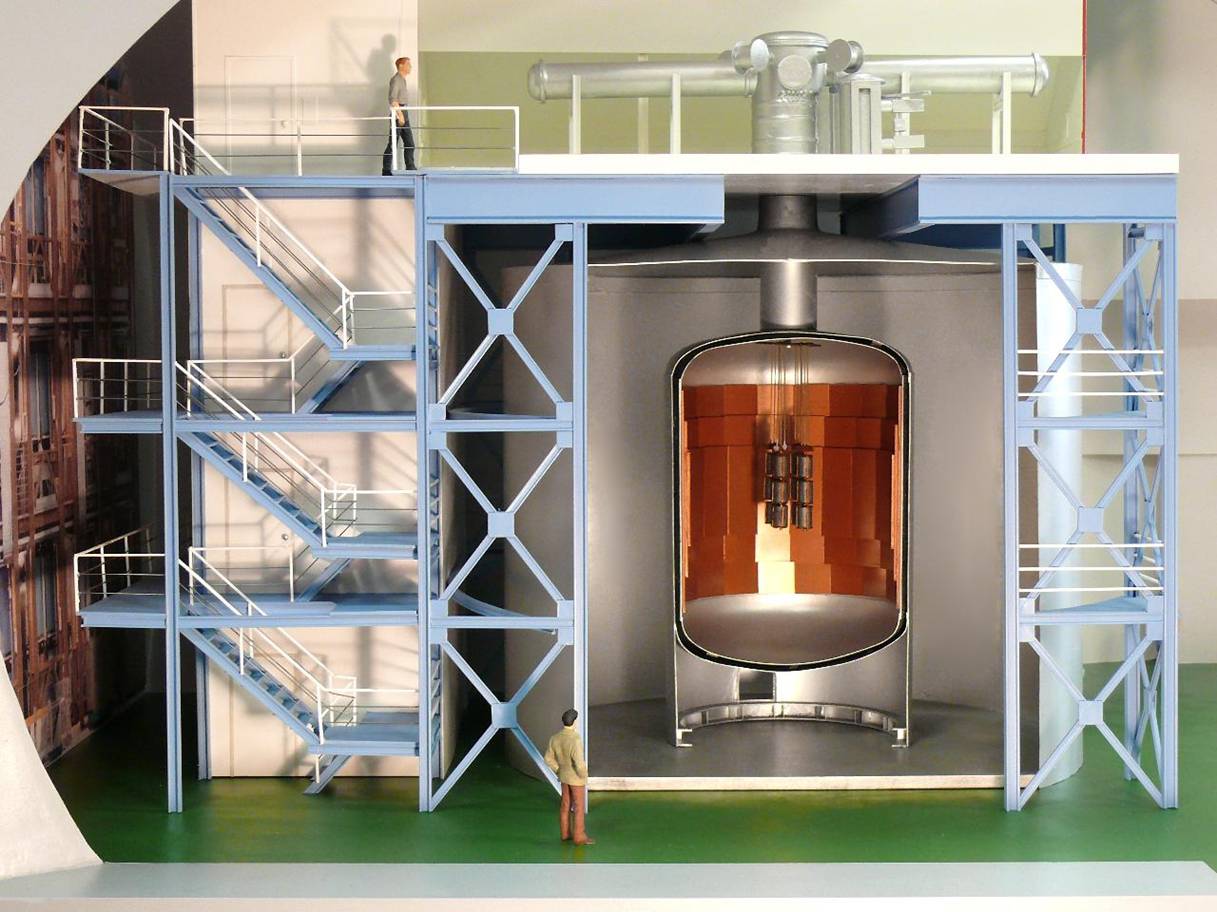}
\includegraphics[height=.2\textheight]{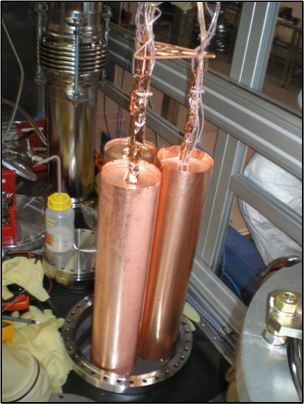}
\includegraphics[height=.2\textheight]{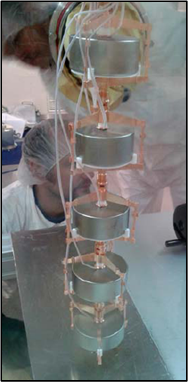}
  \caption{\label{fig:gerda_setup}Left: Schematic model of the \GERDA\, experiment. Center: Three strings with three coaxial detectors each surrounded by the mini-shroud. Right: String with five BEGe detectors.}
\end{figure}

Observation of neutrinoless double beta ($0\nu\beta\beta$) decay would be the first measurement of a 
Lepton number violating process. The decay can only occur, if neutrinos have Majorana character, i.e. they are identical with their own antiparticles.
$0\nu\beta\beta$-decay can be induced by Lepton number violating, beyond standard model processes.
One likely mechanism is the exchange of a light Majorana neutrino. If this is the dominant process leading to $0\nu\beta\beta$ decay, information on the absolute mass scale of neutrinos can be derived. In this case the decay rate depends on the neutrino mass,
\begin{equation}
{T_{1/2}}^{-1} = G(Q^5,Z) \cdot {{\cal M}_{0\nu}}^2  \cdot {\langle\,m_{ee}\,\rangle}^{2}
\end{equation}
where $G$ is the phase space factor, $Q$ is the Q-value of the decay (2039\,keV for $^{76}$Ge), ${\cal M}_{0\nu}$ is the matrix element for the decay and $\langle\,m_{ee}\,\rangle$ the effective Majorana neutrino mass.
The next generation of experiments will reach sensitivities to probe  half lifes in the range of 10$^{26}$ years leading to limits on $\langle\,m_{ee}\,\rangle$ of $\approx$\,0.1\,eV. 

The \GERDA\, experiment, located at the Gran Sasso underground laboratory LNGS of the INFN in Italy with 3400\,mwe, uses High Purity Germanium (HPGe) detectors made from material isotopically enriched to $\approx$87\% in $^{76}$Ge
submerged in liquid Argon (LAr) to search for $0\nu\beta\beta$ decay of $^{76}$Ge \cite{gerda_technical}.

\section{The \GERDA\, experiment}
The main design feature of the \GERDA\, experiment the use of cryogenic liquid Argon (LAr) as the shield against gamma radiation \cite{heu}. HPGe detectors are immersed directly in the LAr which also acts as the cooling medium. The LAr is surrounded by a buffer of ultra-pure water. The water is acting as an additional gamma and neutron shield. It is also instrumented with PMTs to detect through going muons. Fig.~\ref{fig:gerda_setup} shows a schematic model of the setup. 
The detectors are deployed into the LAr in up to seven strings from the top of the cryostat via a lock system mounted inside a clean room. Details of the hardware of the experiment are described elsewhere \cite{gerda_technical}.

The experiment is carried out in two phases. In Phase I coaxial HPGe detectors from the Heidelberg Moscow (HdM) \cite{HdMo} and IGEX \cite{IGEX} experiments were deployed together with up to four HPGe detectors made from natural germanium. Additionally, five newly produced BEGe detectors were deployed \cite{gerda_BEGes}.

In \GERDA\, Phase II 30 new BEGe detectors will be deployed next to the coaxial HPGe detectors already used in Phase I.
LAr scintillates at 128\,nm upon energy deposition by radiation. This will be exploited by a LAr veto system. This allows to identify events with energy deposition simultaneously in HPGe detectors and the surrounding LAr as background. 
The LAr veto consists of a hybrid system, making use of PMTs \cite{large} and SiPMs coupled to scintillating fibers \cite{sipms}. In total 16 3''  Hamamatsu PMTs above and below the detector array will be mounted. Around the HPGe array a curtain of scintillating fibers connected to SiPMs will be used.

\begin{figure}[b!]
\includegraphics[height=.2\textheight]{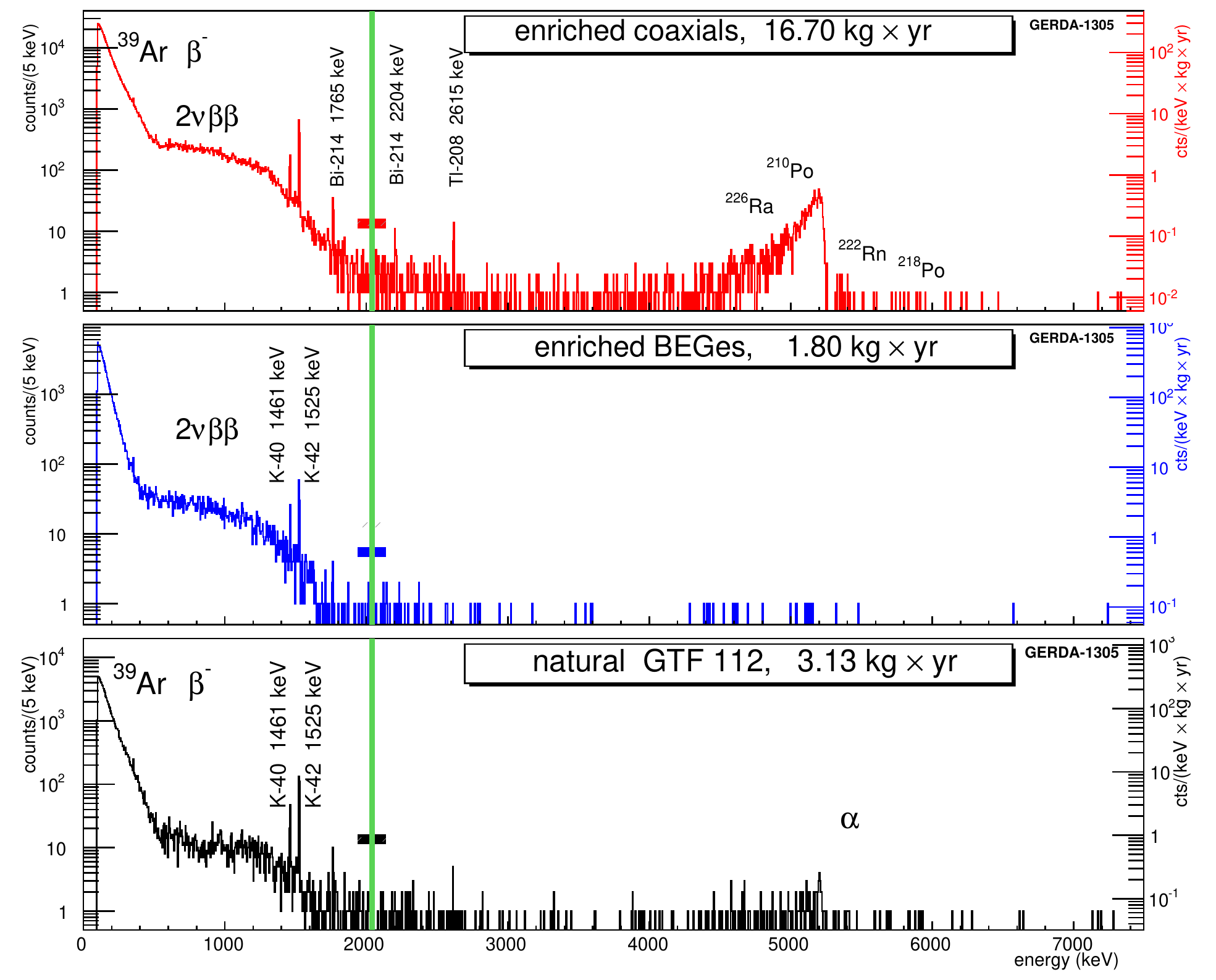}
\includegraphics[height=.2\textheight]{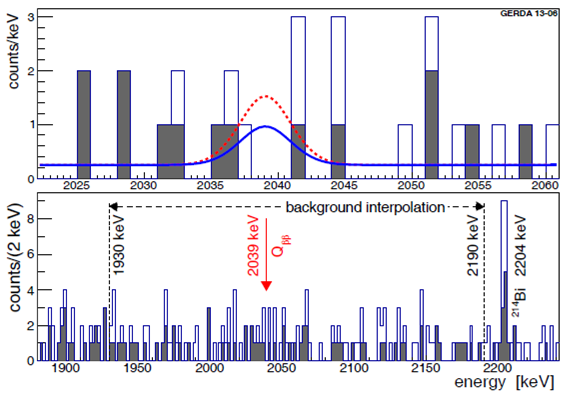}
  \caption{\label{fig:phase1_result} Left: Background spectra of \GERDA\, Phase I for the coaxial enriched detectors (top), the BEGe detectors (center) and the coaxial detectors with natural abundance (bottom) \cite{gerda_background}.  Right: Background spectrum in the energy region of interest No evidence for a peak at 2039\,keV is found \cite{gerda_results}.}
\end{figure}

\section{Phase I of the \GERDA\, experiment}
Pictures of all detector strings before deployment are shown in Fig. \ref{fig:gerda_setup}(center, right). Phase I detectors were surrounded with a copper mini shroud in order to mitigate background from $^{42}$K. This was necessary as the  $^{42}$Ar  ($T_{1/2}$=33\,yr) content in natural LAr is higher then expected. Additionally, the daughter isotope, $^{42}$K ($T_{1/2}$=12.4\,hr), is ionized directly after the decay and can drift in the electric field surrounding the detectors onto the detector surface or to the close surrounding. This leads to an unacceptably high background if no action is taken. The mini shroud drastically reduces the LAr volume from which $^{42}$K ions can be attracted to the detector surfaces. Additionally the field lines are closed by the grounded mini shroud, further reducing the drift of $^{42}$K ions onto the detectors.

Phase I data taking took place from November 2011 until June 2013. For data analysis a total exposure of 21.6\,kg\,yr with coaxial enriched detectors and Phase II BEGe detectors was taken. 

The measured sum spectrum for the enriched coaxial, the BEGe detectors and the coaxial detectors from natural germanium are shown in Fig. \ref{fig:phase1_result} (left) for the energy range 100\,keV to 7.5\,MeV. Also shown is the measured total spectrum in the energy region of interest (RoI) (right). In the latter the total spectrum is displayed with open, while those events surviving a pulse shape cut based on artificial neural networks \cite{gerda_pulse_shape} are displayed by the closed histogram. 
No evidence for a peak at the Q-value of $^{76}$Ge was found. A frequentist profile likelihood analysis was performed. A lower limit of \thalfzero\,$>$ 2.1$\cdot$10$^{25}$\,yr with a median sensitivity of 2.4$\cdot$10$^{25}$ yr was derived \cite{gerda_results}. The expected signal with a half life according to the limit is also displayed as the blue line. This result strongly disfavors the positive evidence for observation of  0$\nu\beta\beta$ decay by parts of the HdM collaboration \cite{evidence}.

\begin{figure}
\includegraphics[height=.2\textheight]{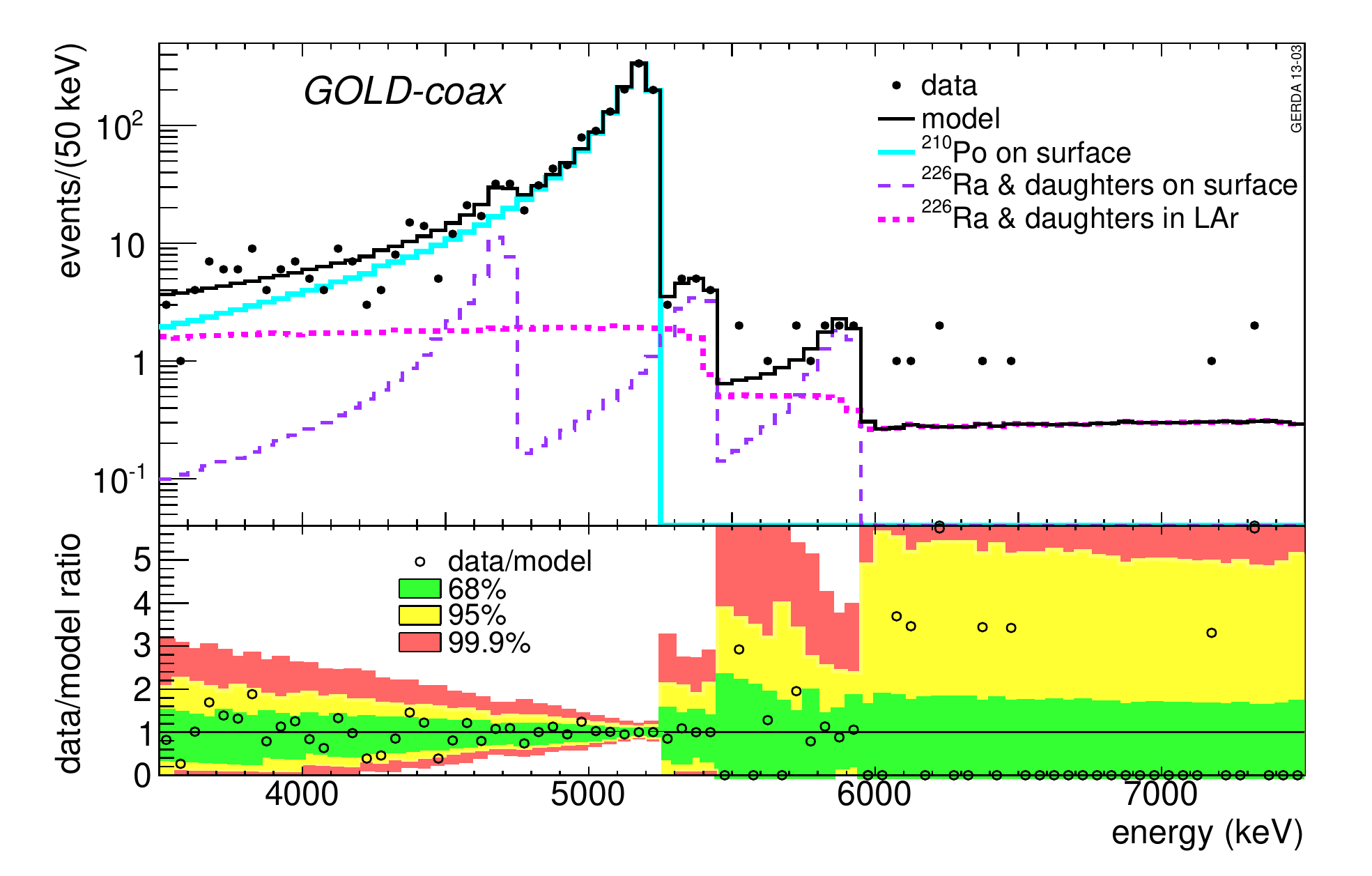}
\includegraphics[height=.2\textheight]{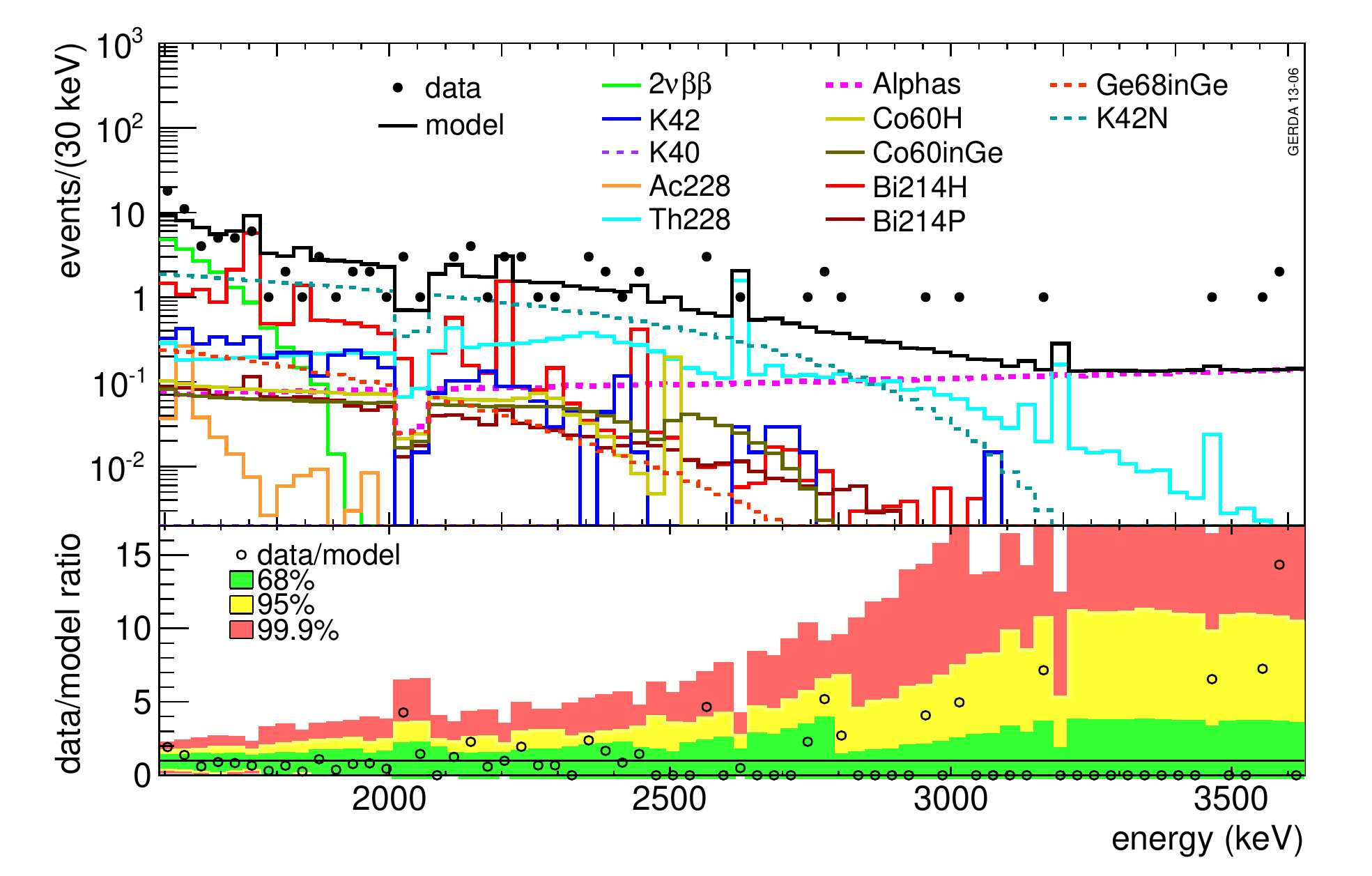}
  \caption{\label{fig:bkg_model} Left: Background decomposition in the energy region 3 -- 7.5\,MeV for the coaxial detectors. Right: Background decomposition for BEGe detectors in the energy range 1.5 -- 3.5\,MeV. Figures taken from \cite{gerda_background}.}
\end{figure}

A background model was derived using knowledge from screening measurements of the used components \cite{gerda_technical} and from the visible structures and peaks observed in the measured spectra shown in Fig. \ref{fig:phase1_result} \cite{gerda_background}.
Using a minimum model accounting only for background components resulting from material contaminations in the close vicinity of the detectors ($<$\,2\,cm) describes the measured background spectra well. According to this model the background in the RoI for the coax detectors results from $^{228}$Th, $^{60}$Co and $^{214}$Bi in the detector assembly, from $^{214}$Bi on the p+ surface of the detectors, from $^{42}$K in the LAr. Additionally, cosmogenically produced $^{60}$Co in the germanium contributes significantly. The structure of the spectrum above 3\,MeV shown in Fig. \ref{fig:bkg_model} (left) for the coaxial detectors can be explained by contamination of the coaxial detector p+ surfaces with $^{210}$Po and $^{226}$Ra daughters with additional contribution from $^{226}$Ra daughters in the LAr directly surrounding the detectors, also shown in Fig \ref{fig:bkg_model}. The latter component does not lead to peak-like structures, rather to a flat distribution extending also to lower energies. This contribution leads to significant background also in the RoI. For the BEGe detectors an additionally component from $^{42}$K on the n+ surface has to be taken into account. Indeed this is the dominant component in the RoI for BEGes, while for coax detectors this contribution is negligible. The difference is explained by the different n+ dead layer thicknesses of BEGes and coaxial detectors. Fig. \ref{fig:bkg_model} (right) shows the background decomposition for the BEGe detectors. The background contributions are summarized in Tab. \ref{tab:backgrounds}.

\begin{table}[b!]
\begin{tabular}{ll|rc|rc|c}
\hline
Isotope & location & \multicolumn{2}{c|}{Coax} &\multicolumn{2}{c|}{BEGe} & Phase I expectation\\
        &          &  Best fit & Range        &Best fit & Range & after PSD \& LAr veto\\
\hline
        &          &  \multicolumn{5}{c}{[10$^{-3}$cts/(keV kg yr)]}\\

\hline
 $^{42}$K & LAr homogeneous & 3.0 & 2.9--3.1 & 2.0 & 1.8 -- 2.3  &  $<$0.1\\
 $^{42}$K & n+ surface      &      &         & 20.8 & 6.8--23.7  & $\approx$0.9 \\
 $^{60}$Co &det assembly    & 1.4  & 0.9--2.1 &     & $<$4.7     &  $<$0.1\\
 $^{60}$Co &Germanium$^{*}$ 
                           & 0.6  & 0.1--0.6 & 1.0   & 0.3--1.0 &  $<$0.1\\
 $^{214}$Bi &det assembly    & 5.2  & 4.7--5.9 & 5.1  & 3.1--6.9 & $\approx$0.3\\
 $^{214}$Bi &p+ surface $^{+}$
                            & 1.4  & 1.0--1.8 & 0.7    &  0.1--1.3&   $<$0.1\\
 $^{228}$Th &det assembly    & 4.5  & 3.9--5.4 & 4.2  & 1.8--8.4  &  $<$0.1\\
 $\alpha$ model & p+ surface and & 2.4 & 2.4--2.5&1.5 & 1.2--1.8 &  $<$0.1\\
                & LAr close to p+&     &&&&\\
\hline
Total           &           & 18.5 & 17.6--19.3 & 38.1 & 32.2--43.3 & $\approx$1.2\\
\hline
\multicolumn{7}{l}{$^{*}$upper limit due to known exposure to cosmic rays above ground \hspace{0.75cm} $^{+}$prior deduced from alpha model}
\end{tabular}
\caption{\label{tab:backgrounds} Contaminations contributing to the background in the RoI according to the minimum model.}

\end{table}
A maximum model, allowing for background components due to contamination of materials further then 2\,cm from the detectors does not fit significantly better. Details about the background model can be found in \cite{gerda_background}.
Without pulse shape discrimination (PSD) a background index (BI) of 18.5$^{+0.8}_{-0.9}\cdot$10$^{-3}$\,cts/(keV kg yr) and  38.1$^{+5.2}_{-5.9}$$\cdot$ 10$^{-3}$\,cts/(keV kg yr) was achieved in the RoI for the coaxial and the BEGe detectors, respectively.
With PSD the BI in the RoI for the total data set used for the $0\nu\beta\beta$ analysis  was (1.0$\pm$0.1)$\cdot$10$^{-3}$\,cts/(keV kg yr) \cite{gerda_pulse_shape}.
Especially for the BEGe detectors the PSD efficiency is of high relevance, as the expected background from  $^{42}$K on the n+ surfaces would otherwise deteriorate the sensitivity (see below). 


Using the background model a new value for the half life of 2$\nu\beta\beta$ decay of $^{76}$Ge with significantly lower uncertainties was derived \cite{gerda_majoron}. The obtained half life of \thalftwo\,=\,(1.926$^{+0.025}_{-0.022}$(stat)$\pm$0.092(syst))$\cdot$10$^{21}$yr is in good agreement with the value obtained in an earlier publication with much lower exposure \cite{gerda_2nbb}. Note that within statistical uncertainties there is no significant difference in the extracted half life for different background models assumed. 

Lower limits on the half life of Majoron accompanied $0\nu\beta\beta$ decay have also be obtaied. These are currently the most stringent limits for $^{76}$Ge. For decays with spectral index $n=1$,$ n=3$ and $n=7$ the limits have been improved by a factor of five to six with respect to earlier publications \cite{gerda_majoron}. 
 For a spectral index $n=2$ a limit has been published for the first time.
All half life limits lie between 0.3 and 4.2 $\cdot$10$^{23}$\,yr.

A dedicated analysis based on time correlation of subsequent $\alpha$ decays of the   $^{226}$Ra, $^{228}$Th and $^{227}$Ac decay chains has revealed no candidate events in the full exposure. This allows to set an upper limit for the intrinsic contamination of the high purity germanium with these isotopes of A($^{226}$Ra, $^{228}$Th and $^{227}$Ac)$<$4\,nBq/kg.

\begin{figure}
\includegraphics[height=.15\textheight]{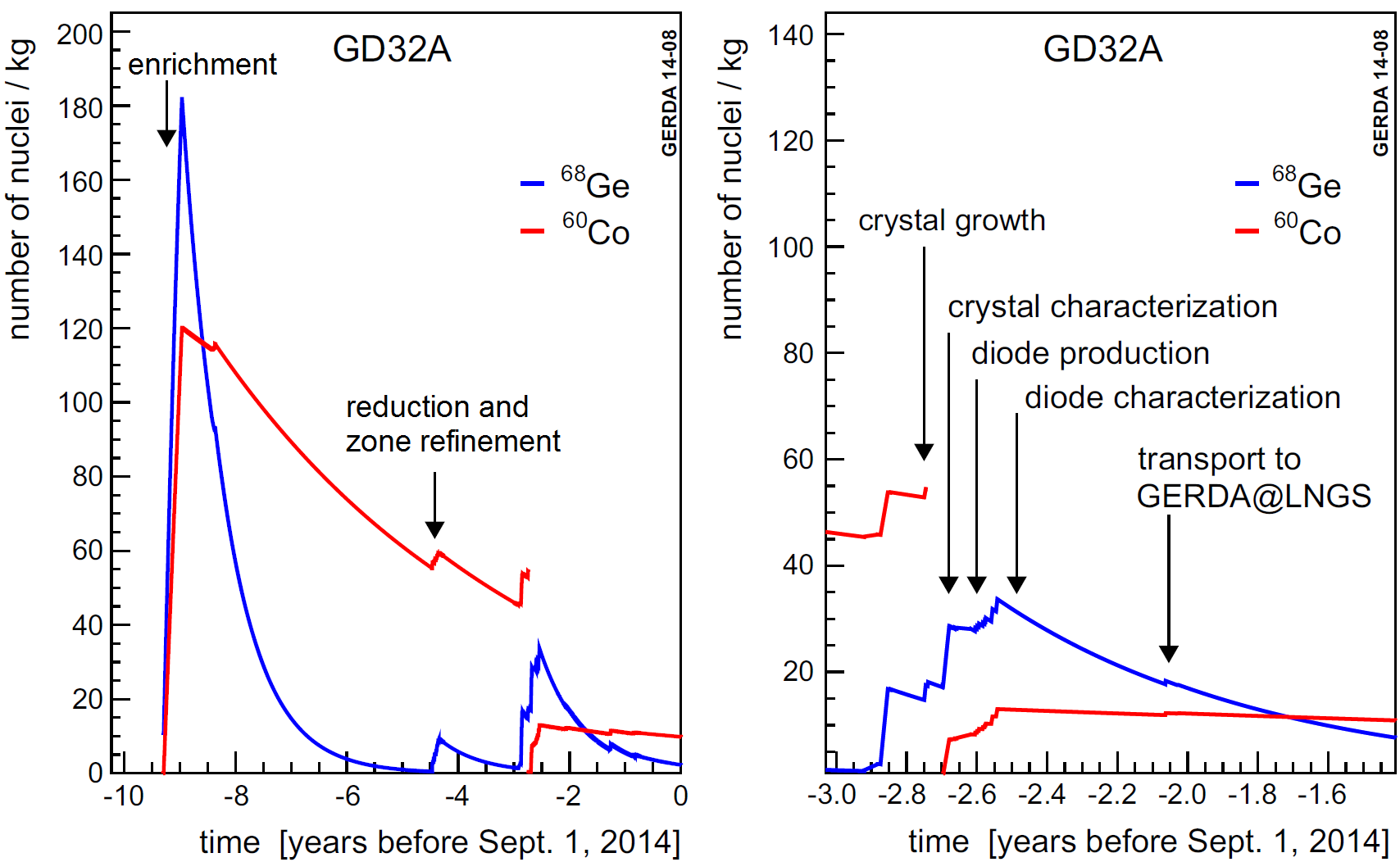}
\includegraphics[height=.15\textheight]{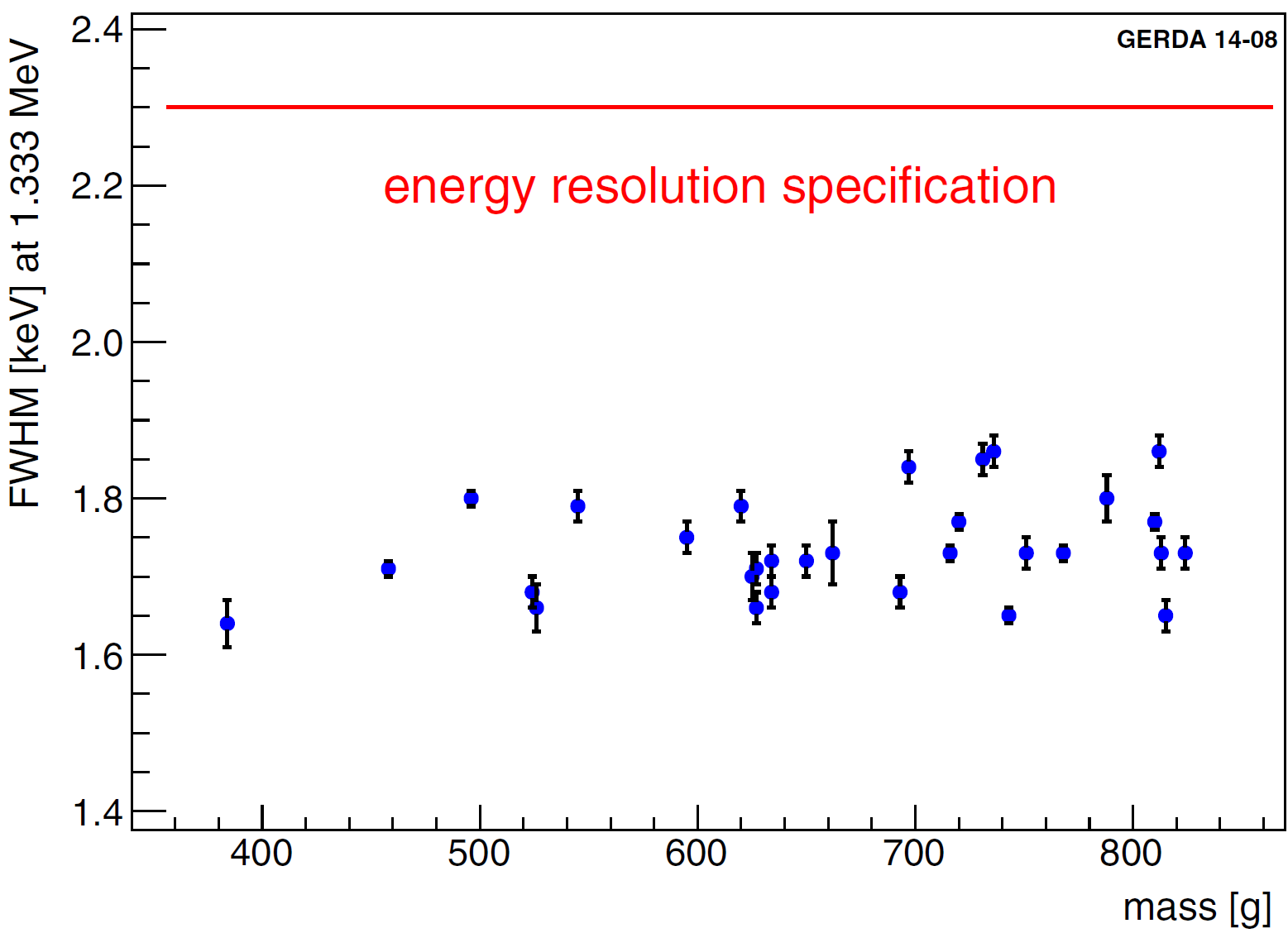}
  \caption{\label{fig:bege_history} Left and center: Activation history of one of the BEGe detectors. Right: Energy resolution of all 30 BEGe detectors as a function of total mass. Figures taken from \cite{gerda_BEGes}}
\end{figure}

\section{Preparations for Phase II}
In the second phase of the \GERDA\, experiment 30 BEGe detectors will be deployed together with the Phase I coaxial detectors.
A total mass of 35\,kg germanium, enriched to $\approx$87\,\% will be available. BEGe detectors have an increased PSD capability due to their special point like electrode configuration \cite{PPC, BEGes}. 

35.5\,kg of GeO$_{2}$ were procured by ECP in Siberia, Russia. This material was reduced to germanium metal ingots and zone refined to 6N electronic grade purity at PPM in Lingoslheim, Germany. Final zone refinement and crystal pulling happened at Canberra in Oak Ridge, US, while diode production was done at Canberra in Olen, Belgium. 

\begin{figure}[b!]
\includegraphics[height=.175\textheight]{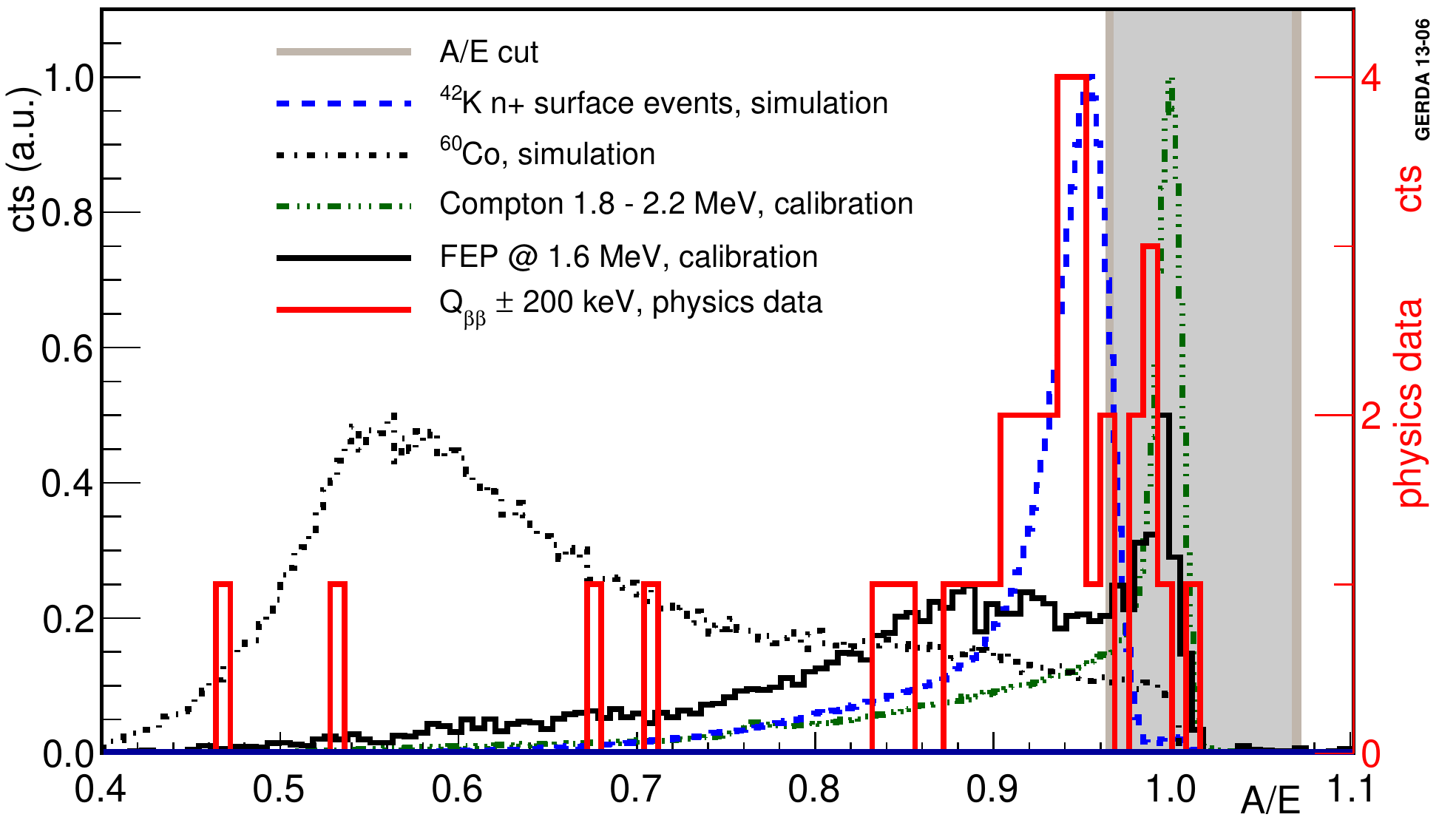}
  \caption{\label{fig:ae}  A/E distribution of BEGe detector events in the energy range $\pm$100\,keV around 2040\,keV   from Phase I data.}
\end{figure}

During all times crucial care was taken to properly shield the material against cosmic ray exposure whenever possible. Transport of the enriched germanium was mainly performed in a specially designed container consisting of a steel shield with 0.7\,m top thickness and an additional 0.7\,m thick water shield. 
The container was transported bewtween US and Europe in the deapest part of the transport ship.
During processing, whenever the germanium was not needed. it was stored in shallow depth underground sites. Exposure to cosmic rays was tracked in a dedicated data base. The expected activation history in terms of $^{68}$Ge and $^{60}$Co concentration  due to exposure to cosmic rays  for the example of one of the diodes is displayed in Fig. \ref{fig:bege_history}. The individual processing, transportation and storage steps can be clearly seen. Tracking the exposure history allowed to make predictions for the $^{68}$Ge and  $^{60}$Co concentrations for each individual detector. For June 2015 in total 13.9 and 23.7 $^{68}$Ge and  $^{60}$Co nuclei are expected per kg of detector material, respectively \cite{gerda_BEGes}.

All 30 BEGe detectors were subsequently characterized in a vacuum cryostat in the HADES underground lab in Mol, Belgium.  The energy resolution (FWHM at 1.3\.MeV) of each enriched BEGe is shown as a function of the total detectors mass in Fig. \ref{fig:bege_history}. All 30 BEGe detectors had energy resolutions between 1.65\,keV and 1.85\,keV, significantly better then specified  with Canberra before detector production started ($<$2.3\,keV at 1.3\,MeV) \cite{gerda_BEGes}.

\begin{figure}[t!]
\includegraphics[height=.15\textheight]{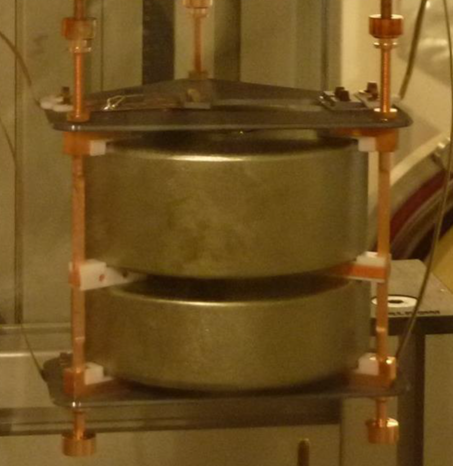}
\includegraphics[height=.15\textheight]{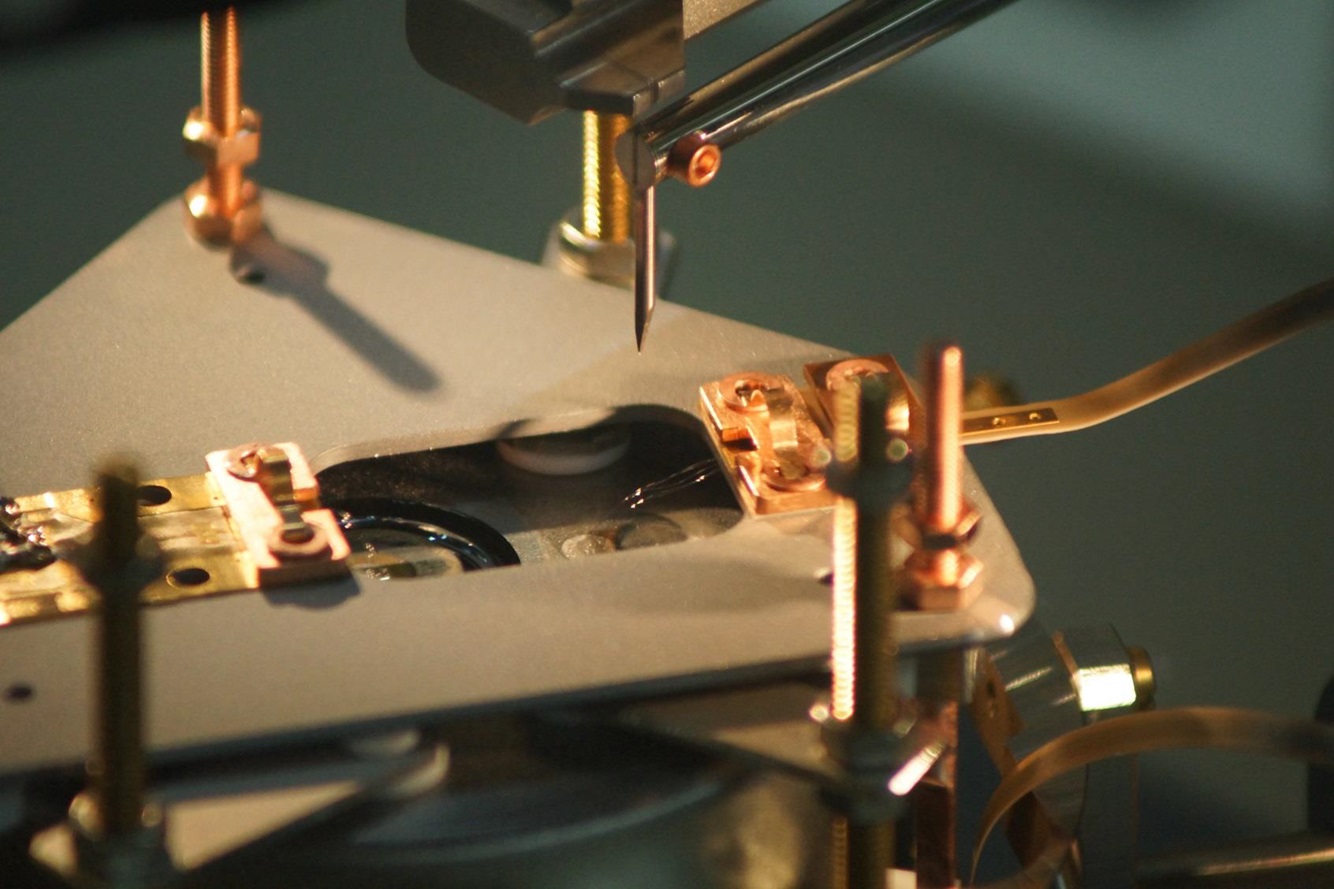}
  \caption{\label{fig:bege_holder} Left: BEGe pair mounted in their holder. Right: Bonding of BEGe contacts.}
\end{figure}

\begin{figure}[b!]
\includegraphics[height=.175\textheight]{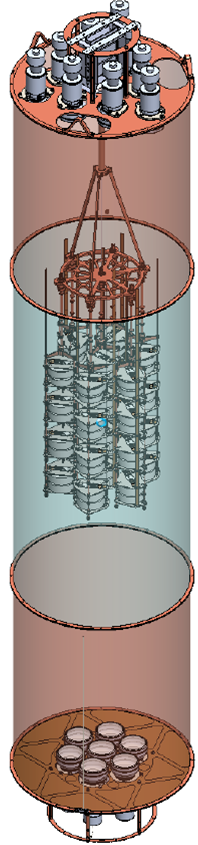}
\includegraphics[height=.175\textheight]{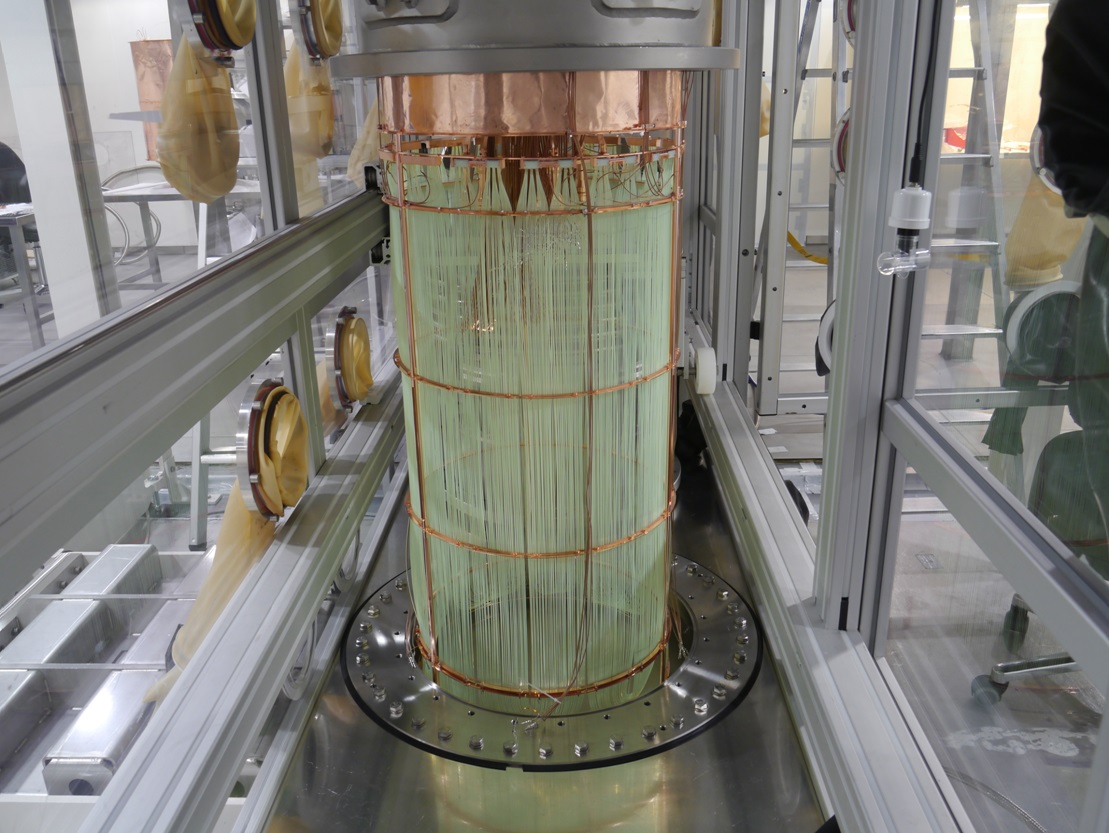}
\includegraphics[height=.175\textheight]{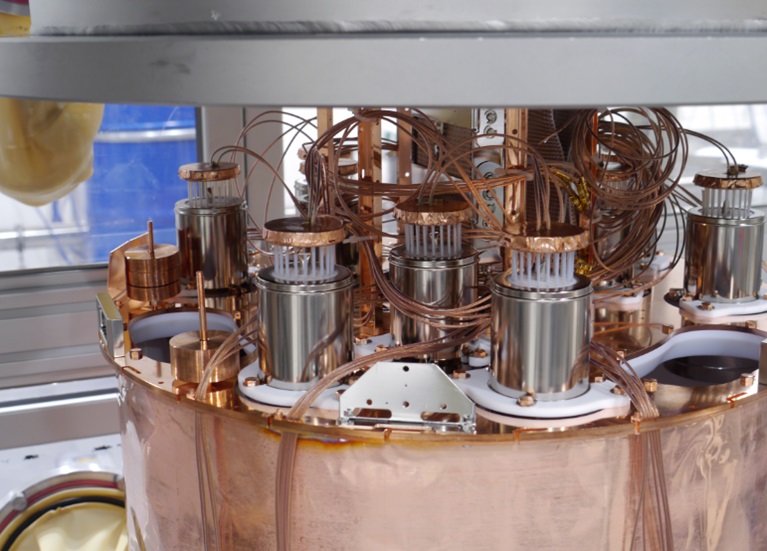}
  \caption{\label{fig:lar_veto} 
Left: Drawing of the whole detection system containing the HPGe detector array in the center. 
Center: It is surrounded by a curtain of fibers that are connected to SiPM detectors. Right: Above and below the detector array PMTs are mounted.}
\end{figure}

Fig. \ref{ae}  shows the distribution of the A/E parameter \cite{gerda_pulse_shape} for all BEGe events in the $\pm$100\,keV energy range around 2039\,keV (excluding events in the energy range $\pm$10\,keV around 2039\,keV, as these were blinded when the analysis was made) recorded in Phase I background data together with expected distributions from measurements in a test facility and simulated distributions. It can clearly be seen that the bulk of the events follow the A/E distribution expected for events resulting from the decays of  $^{42}$K on the n+ surfaces and can be discriminated with high efficiency from bulk single interaction events. 

For Phase II new detector holders have been designed and produced. Two BEGe detectors will be mounted as a pair in one holder. The holder plates consist of crystalline high purity silicon. Screening measurements have revealed upper limits of 0.2 mBq/kg for $^{226}$Ra and 0.15 mBq/kg for $^{228}$Th (90\% C.L.) contaminations of the used silicon. The plates are connected with bars made from the same screened high purity copper also used for Phase I holders.  Fig. \ref{fig:bege_holder} (left) shows a BEGe pair mounted in their holder.  
Signal transmission and HV supply will happen via flexible flat cables produced from CuFlon (trademark by Polyflon company). 
Detectors in Phase II will be connected to the cables by wires bonded to an aluminum pad evaporated onto the detector n+ and p+ surfaces on the one side and to a bond pad on the signal and HV cables on the other.
 Fig. \ref{fig:bege_holder} (right) shows a detector in its holder inside its mounting jig during bonding of the contact wires.

A new lock system, allowing to deploy all Phase II and Phase I detectors, has been constructed and mounted on top of the \GERDA\, cryostat. 
The LAr veto has been assembled and extensively tested before installation into the \GERDA\, setup. 
Fig. \ref{fig:lar_veto} (left) shows a schematic drawing of the whole detection system. 
Fig. \ref{fig:lar_veto} (center, right)  shows pictures of the fiber shroud around the detector array and the top PMT plate as mounted in the \GERDA\, experiment, respectively.  The detector strings will be surrounded by a mini shroud  to protect the detectors against $^{42}$K ions drifting onto their surface. It is made from TPB coted low background nylon used also for the Borexino experiment \cite{borexino}.

\begin{figure}
\includegraphics[height=.2\textheight]{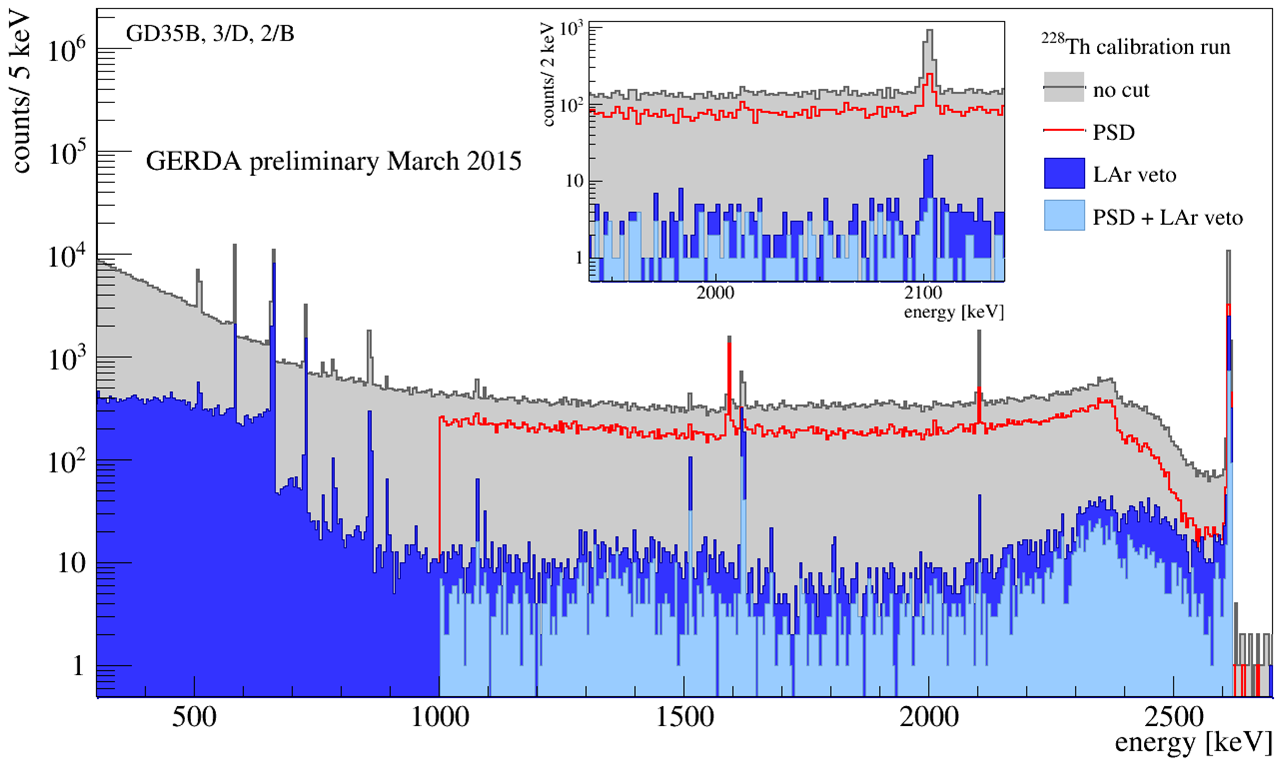}
  \caption{\label{fig:commissioning_data} First commissioning data with the full LAr veto system mounted around a string of BEGe detectors. The spectrum is shown without any cuts (gray), after detector AC (red), after LAr veto (dark blue) and after the PSD cut (light blue).}
\end{figure}

First commissioning data taken with a detector string containing 3 working BEGes with energy resolution of $\approx$\,3\,keV at 2.6\,MeV and the LAr veto system in place using a $^{228}$Th calibration source are shown in Fig. \ref{fig:commissioning_data} (right). The supression factor of events induced by the  $^{228}$Th calibration source in the RoI is $\approx$\,100 also taking into account detector anti coincidence (AC). It is expected that the supression factor will be even higher with the full detector array mounted as the AC efficiency will increase.

With the LAr veto and PSD it is expected that the BI for Phase II will be dominated by 0.9$\cdot$10$^{-3}$ cts/(keV\,kg\,yr) of $^{42}$K decays on the n+ surfaces of the BEGes.  $^{214}$Bi and $^{228}$Th in the close surrounding and in the front end electronics are expected to contribute $\approx$0.3$\cdot$10$^{-3}$ cts/(keV\,kg\,yr) to the BI in the RoI (see Tab.\ref{tab:backgrounds}. All other components are expected to be insignificant. With an exposure of 100\,kg\,yr and a total BI of $\approx$1.2$\cdot$10$^{-3}$ cts/(keV\,kg\,yr) a medium limit setting sensitivity for the half life of $0\nu\beta\beta$ decay of $^{76}$Ge of $\hat{T_{1/2}}\,\approx$\,10$^{26}$yr can be reached.

\section{Conclusion and outlook}
The \GERDA\, experiment has finished Phase I data taking and produced competitive lower limits on the half life of $0\nu\beta\beta$ decay of $^{76}$Ge, \thalfzero\,$>$\,2.1$\cdot$10$^{25}$\,yr. Competitive half life limits on the Majoron accompanied  $0\nu\beta\beta$ decays of $^{76}$Ge and a new value for \thalftwo($^{76}$Ge)\,=(1.93$\pm$0.10)\,yr with reduced systematic uncertainties have been deduced from Phase I data. Hardware and infrastructure for Phase II of the \GERDA\, experiment have been prepared and are being commissioned. First data reveal that the LAr veto works as expected. Data taking will start soon. The expected BI in the RoI is $\approx$10$^{-3}$\,cts/(kg\,keV\,yr). This will allow to reach a limit setting sensitivity of 10$^{26}$\,yr.


\bibliographystyle{aipproc}


\end{document}